# Shifting Mobility Behaviors in Unprecedented Times: Intentions to Use On-demand Ride Services During the COVID-19 Pandemic

**Maher Said**
(corresponding author)
Department of Civil and Environmental Engineering
Northwestern University
A308 Technological Institute
2145 Sheridan Road, Evanston, IL, 60208, USA
Email: MaherSaid@u.northwestern.edu

**Jason Soria**
Department of Civil and Environmental Engineering
Northwestern University
A308 Technological Institute
2145 Sheridan Road, Evanston, IL, 60208, USA
Email: jason.soria@u.northwestern.edu

**Amanda Stathopoulos**
Department of Civil and Environmental Engineering
Northwestern University
A312 Technological Institute
Phone: +1 847-491-5629, Fax: +1 847-491-4011
2145 Sheridan Road, Evanston, IL, 60208, USA
Email: a-stathopoulos@northwestern.edu

**ABSTRACT**

The spread of COVID-19 has been a major disruptive force in people's everyday lives and mobility behavior. The demand for on-demand ride services, such as taxis and ridehailing has been specifically impacted, given both restrictions in service operations and user's concerns about virus transmission in shared vehicles. During the pandemic, demand for these modes have decreased by as much as 80%. This study examines intentions to use on-demand ride services in a period of drastic changes in lifestyles and daily routines coupled with unprecedented mobility reductions. Specifically, we examine the determinants for the shift of intentions to use these on-demand modes of travel in the early stages of the pandemic. Using data from a survey disseminated in June 2020 to 700 respondents from contiguous United States, ordinal regression modeling is applied to analyze the shift in consideration. The results indicate that political orientation and health-related experiences during the pandemic are significant sources of variation for individual changes in intentions to use ridehailing. Additionally, characteristics such as age and income result in consideration shifts that contradict the typical ridership profiles found in the ridehailing literature. Specifically, on-demand ride consideration decreases as a function of age and income. Moreover, transit-users are more willing to consider on-demand rides than private vehicle users, suggesting that shared vehicle modes have a similar risk-profile. We discuss the role of on-demand ride services in the pandemic era, and the need to investigate political orientation and evolving pandemic experiences to pinpoint their role in future mobility systems.

## INTRODUCTION

Sustainable urban transport increasingly relies on mobility-as-a-service models that bundle mobility options (*1*). Shared and on-demand modes like transit, bike-sharing and ridehailing are central components in these systems to offer travelers a sustainable and competitive alternative to car use and ownership (*2-4*). Ridehailing mobility platforms (such as Uber, Lyft and BlaBlaCar) in particular have resulted in rapid and profound shifts in mobility systems (*5*). Since the inception of Uber in 2010, the familiarity with and role of ridehailing has grown considerably. 97% of Americans are familiar with the concept, while 36% have used ridehailing services (*6*). Even with the rapid diffusion of these on-demand ride platforms, it is difficult to pinpoint the precise impact on urban mobility systems. On one hand, ridehailing promises to offer on-demand flexibility for customers, improve first-last mile connections to transit, and potentially increase vehicle occupancy (via pooling) thereby reducing vehicle miles and congestion (*7-12*). Other work suggests that ridehailing services compete with public transit, induce new travel, increase congestion during peak periods, and mislead consumers and drivers through opaque pricing and labor practices (*13-16*). An important reason that the impact and role of ridehailing platforms has remained uncertain is the significant variation observed across cities, service-types and user segments. Evidence from multi-city studies suggests that driving is responsible for the greatest modal loss towards ridehailing (*7, 17*). Yet, more recent work examining trip-level data finds ridehailing to be the biggest contributor to growing traffic congestion in San Francisco (*18*). The specific demand relationships likely depend on several factors such as urban density, type of transit-TNC (transportation network company) competition, as well as demographic composition and local car-ownership (e.g. (*19-21*)).

This research summary highlights that the advent of innovative mobility services offering short-term access to transportation on an as-needed basis is changing how travelers move in cities. These changes call for policymakers to rethink the design and access to mobility, accounting for the emerging travel behavior as well as changing mode relationships, ownership and access patterns. Adding to the challenge is the fact that ridehailing services are constantly evolving, with operators launching novel options like ridepooling where riders split the ride and cost. These new service offerings lead to new demand behaviors, and attention to new behavioral factors such as privacy concerns or sharing attitudes towards pooled rides (*12, 22, 23*).

Over the past year, the COVID-19 pandemic has overhauled everyday travel patterns, causing major disruptions to the use of shared vehicle services, represented by transit and on-demand ride services. With mandated lockdowns, limitations for non-essential out-of-home activities and wide-spread social distancing, demand for shared vehicle mobility modes has fallen considerably, with reductions in demand reaching 90% at the height of the pandemic (*24-27*). The major pandemic related mobility-disruptions observed during 2020 calls for a renewed and urgent need to examine the demand behavior of on-demand ride service users.

The reduction in shared mobility is due in part to supply reduction, service closures and recommendations to engage in social distancing and reduce non-essential travel (*26*). Nevertheless, there is significant variation of demand, across user segment and geographies calling for a more thorough understanding of voluntary mobility use reductions. A growing body of recent works has revealed systematic trends in transit reduction patterns during COVID-19 in the United States (*28-30*). Yet, what is less well understood is the variation and motivation for different levels of user acceptance behaviors. Moreover, outside of transit, there is still limited understanding of how other shared vehicle transportation is perceived and affected.

The goal of this study is to examine the acceptability of using on-demand rides, like ridehailing and taxi services during the COVID-19 pandemic. The continued use and recovery of demand for on-demand ride services like ridehailing and taxis, has important societal implications as part of a broader mobility portfolio and as an alternative to private vehicle use and ownership (*31*). This research investigates the variation in use intentions according to different population segments, travel habits, experiences during the pandemic, and perceptions. Specifically, we analyze the variation across population segments and pinpoint the role of unique pandemic-era factors, namely adverse household experiences from COVID-19 and political orientation that has been shown to affect pandemic behaviors (*32*). We hypothesize that along with demographic and attitudinal variables, previous exposure to the virus along with severity of symptoms can negatively affect willingness to use shared transportation. Moreover, given the significant polarization in the U.S. surrounding pandemic behavior, we examine the role of political orientation on pandemic risk-perceptions and thereby use intentions. Understanding how different population segments approach the decision about using on demand ride services in pandemic times, along with their main motivations and concerns, is critical to build fundamental insight on the adaptation of preferences for different modes during a viral pandemic outbreak. The modelling is based on 2020 web survey data from 700 U.S. residents analyzing the willingness to use on demand rides prior to and during the pandemic.

The remainder of this paper is organized as follows. The following section presents a literature review on hailed ride services as well as information on the COVID-19 pandemic and its impact on travel and mobility. Next, the data collection process is described, and preliminary insights are provided. The fourth section presents the modeling approach, followed by subsequent results and a discussion in the next section. Finally, the paper is then concluded with remarks on limitations and future research.

## LITERATURE
### Ridehailing demand state of practice
Before the implementation of major restrictions on mobility and wide-spread health concerns related to COVID-19 spread in the general public, ridehailing had begun playing an important role in urban transportation systems. Prior to the pandemic, transportation network companies (TNCs) provided millions of daily rides with large markets, such as Chicago, completing nearly 300,000 daily trips (*33*). Research suggests that ridehailing use is higher in younger, well-educated and higher income population segments (*13, 17, 34-36*), with most trips taking place in highly dense areas for diverse purposes, mainly related to social or leisure travel (*13, 37*). In terms of attitudinal drivers, TNC users also tend to have variety seeking and technology embracing attitudes (*12, 38*). Research has also found that some car users switch to ridehailing for its convenience and comfort (*39*).

On the whole, the research highlights both advantages and challenges related to increasing use of ridehailing platforms. Benefits include serving as a gap filler for public transit, providing personalized door-to-door mobility that is relatively affordable compared to traditional taxis, and optimizing trip trajectories for pooled rides (*8, 9, 40*). Additionally, there is evidence of social equity benefits where ridehailing pickups are less discriminating against minority neighborhoods than traditional taxis (*41*). At the same time, there is a growing body of evidence concerning negative externalities related to vehicle mile increases and cannibalization of transit ridership. Ridehailing use has been linked to increases in congestion and vehicle-miles traveled via deadheading (*18, 42, 43*). Concerning public transit ridership relationships, several research studies show mixed results on whether TNCs complement (*19, 44*) or competes with transit (*21*), and findings appear to vary according to vehicle ownership and use frequency (*43*), and availability of other transportation system options (*20*).

### Potential for pooled rides
Simulation analysis has suggested that pooling of rides can achieve significant gains in efficiency with reasonable delays (*8*). Yet, the current empirical work finds that demand for pooling is typically too modest to produce significant system VMT savings, with pooling rates between 6% and 35% (*21, 33, 37, 45-47*). Additionally, analysis shows that 94% of ridepooling trips are made by just 10% of riders (*48*). The potential for significant gains is therefore limited by the narrow set of users willing to pool rides. Presently, lower-income and denser areas see higher rates of sharing (*48, 49*). Moreover, in terms of user priorities, it appears that time-cost trade-offs are more important to users than the sharing aspect (*22*). Notably, during the COVID-19 pandemic, many jurisdictions have suspended the use of pooling to prioritize the health of drivers and users, thereby limiting the ability to empirically study and understand motivations of users opting for on-demand rides and pooling. In our survey-based research we are able to analyze the acceptability of on-demand rides by exploring the stated intentions of users.

**Taxi-Ridehailing demand relationship**
Prior to the impact of COVID-19 on the transportation industry, traditional taxi services have been experiencing a steady decline in ridership over the years as a result of intense competition. Demand substitution tends to be higher between taxi and ridehailing services than for other modes (*31, 35, 50*). As a result of stringent regulations for operations and medallions (*5*), and competition from ridehailing drivers, taxi demand on the whole has seen a steep decline. Increasingly, taxi service providers promote similar service-models to ridehailing, including on-demand and curb-hailing services as well as online payments. Still some differences remain. Taxis are subject to stricter regulations, have flat pricing, and typically complete transactions at the end of a ride. Ridehailing, on the other hand, provides door-to-door services, uses dynamic pricing, completes all transactions through smartphones, and is less regulated (*51, 52*). While some service and user differences exist, for example a preference for taxi among older riders, (*53*), in this paper we jointly study the consideration of on-demand rides, encompassing both taxi and ridehailing, to examine COVID-19 era mobility behaviors. We use the term 'hailed rides' to refer to the two services jointly throughout this paper.

**Mobility behaviors in the COVID-19 era: shared modes**
As a result of the critical levels of community spread of the COVID-19 virus, all 50 U.S. states have taken measures to prevent social contact and slow down the diffusion of the virus. The implemented stay-at-home orders and social distancing measures have led to drastic reductions in mobility (*54*). While these measures have affected the entire U.S. mobility system, the reductions in travel have disproportionally affected shared modes. Public transit is especially hard hit with ridership declining by roughly 90% for some systems and has transit officials growing concerned that this may lead to drastic cuts in future service (*30, 55*). Moreover, impacts have been unequal across socio-economic groups, while many workers can work from home, others have jobs that demand physical presence and continued travel to work (*56, 57*). Iio et al. (*29*) point to disparities in the levels of mobility adaptation across income groups in the US during the pandemic, with evidence that the remaining public transit users are more likely to be transit dependent riders with few mobility options.

**Impact on ridehailing and taxi demand**

The ridehailing industry has also been severely impacted. At the onset of social distancing and lockdown policies, ridehailing ridership in North America plummeted by approximately 80% (*25*). Because ridehailing has been classified as an essential service in the United States, the rides they provide during the pandemic are assumed to be limited to essential travel (*58*). In addition to many establishments being closed due to lockdown mandates, this classification severely restricted ridehailing trips for the purpose of accessing leisure and recreation. Among the first actions that ridehailing operators took was to shut down their ridepooling services (e.g. *UberPool* and *Lyft Line*) (*59*). Other operational strategies during the pandemic have been to provide cleaning products, face masks, physical partitions between driver and passengers, and financial support to drivers. As of March 2020, TNCs mandated face masks for passengers in the United States, even though mask wearing was not a federally enforced policy and some states have opted to not required their use. Taxi services have taken similar restrictive measures, with reports suggesting major challenges for the industry and drivers. For example, in Chicago, the number of active medallions has decreased by roughly 80% with some companies reducing their active fleets by as much as 98% in the absence of demand (*24*).

**Motivations of users**

The mobility behaviors emerging during the pandemic in the United States are complex and varies significantly among socio-economic segments (*60*). Specifically, the behaviors related to social distancing measures, such as mask-wearing and spacing in shared vehicles, has become an unexpectedly political issue (*32*). In the U.S. there is a stark partisan divide in the perceived severity of the outbreak and the need to adhere to strict protective measures, both during lockdown and reopening phases (*61-63*). Moreover, the myriad of measures and guidelines put forth by different U.S. states and local jurisdictions has muddled the messaging on interventions and understanding of behavior. It is likely that the politicization surrounding social distancing will continue to shape travel behavior and affect the willingness to use on-demand rides and transit into the future.

While recovery and the full reopening of the economy remains a global goal, the process is challenging, with travel behavior receiving significant attention. Travel choices by individuals, notably out-of-home travel frequency and mode choices, are at the core of determining the effectiveness of lock-down and social distancing measures. Individual travel behavior arises from navigating challenging trade-offs between maintaining livelihoods and mental health via mobility, against compliance and risk-reduction related to COVID-19 exposure via immobility (*56, 64*). Specific evidence for on-demand ride behaviors is scarce. In a Spanish study conducted during the pandemic, survey-takers indicated that only 40.7% would consider using ridehailing even if masks, gloves, and sanitizer gel were provided (*65*). The willingness to pay for ridehailing also declined. Survey-takers in Toronto indicated that they are likely to never use ridehailing again or would rather wait until the virus no longer poses a substantial threat than to use it during the pandemic (*66*).

**Literature takeaways**

In summary, the research shows that shared vehicle mobility behavior during the COVID pandemic, along with the return of non-essential travel, and the future of on-demand ride demand are complex processes affected by both tangible actions (e.g. mask mandates, vaccination rates, evolving lockdown measures) and latent attitudes and beliefs (e.g. acceptance of shared vehicle rides, political orientation). To date, there are important research gaps related to hailed mobility behaviors that call for careful evaluation of the COVID specific context, changing mode relationships, comfort to share rides with strangers, and the emerging role of partisan differences.

In this paper we study the intention to use on-demand ride mobility during the COVID-19 lockdown period, shaped by numerous social distancing measures. Specifically, we explore the shift in user intentions comparing pre-pandemic and June 2020 ridership. To contribute new insight on pandemic mobility decision-making, we carefully examine new factors relevant to understanding decision-making during the on-going pandemic, such as political views and household adverse experiences with COVID-19 or pandemic preventive measures.

**SURVEY DESIGN & DATA COLLECTION**

Data was collected using a web survey designed on Qualtrics and disseminated through the Prolific platform to 700 mainland U.S. respondents in early June 2020. In the era of COVID-19, much behavioral research has relied on web-based survey and samples to obtain timely data and comply with social distancing restrictions (e.g. (*67*)). Nonetheless, online sampling can result in self-selection or coverage bias, specifically related to internet access (*68*). While research on the validity and quality of online sampling is ongoing, evidence indicates that online samples are more diverse and often comparable in quality to traditional survey samples (*69-71*).

## Survey Design

The main purpose of the survey is to collect information and data relevant to shared ride behavior as well as the impact of the COVID-19 pandemic on travel behavior. While the survey consists of 7 major sections, responses from 3 of these sections are relevant to this research. The central survey section for this study focuses on the likelihood of using on-demand ride services for essential travel purposes during the pandemic. Specifically, respondents are asked the following two questions:

1. "Assume that you need to make a trip for an essential purpose **DURING** the outbreak. How likely are you to request a typical ride-hailing service (such as Uber or Lyft) or taxi service?"
2. "Assume that you needed to make a trip for an essential purpose **BEFORE** the outbreak. How likely were you to request a typical ride-hailing service (such as Uber or Lyft) or taxi service?"

For the remainder of the paper, the latter modes will be jointly referred to as *hailed modes* or *services*. The responses are on a 5-point Likert scale ranging from *very unlikely (1)*, *neither likely nor unlikely (3)*, *very likely (5)*. A 5-point scale was used in place of a 7-point option, to prevent respondent fatigue given the large number of sections in the full survey (*72*). The '*before*' question allows us to establish a usage baselines to analyze the shift in consideration. It is important to point out that these questions emphasize trips made for *essential purposes*, as this distinction is likely to influence responses and have an effect on the model presented in later sections of the paper. The survey does not specify what *essential purpose* trips are, leaving this up to respondents to define according to their lifestyles and state guidelines. Further questions in this section ask respondents about their status as essential worker, the extent to which the pandemic affected their lives, whether someone in their household has been laid-off, quarantined or hospitalized in this period and the duration that their household has been under stay-at-home order, if at all.

The next section collects information on the respondents' latent attitudes towards the environment and technology. Respondents are presented with indicator statements such as "I am willing to switch to active modes of transportation (such as walking or cycling) in order to protect the environment" and "Technology is changing society for the better" then asked to rate these statements on a 5-point Likert scale ranging from *strongly disagree (1)*, to *neither agree nor disagree (3)*, to *strongly agree (5)*.

Finally, respondents are presented with questions related to their socioeconomic status and demographics, such as gender, age, ethnicity, employment status, education, household size and income.

## Sample description and Statistics

For this study, out of 700 responses, 1 response is removed for missing critical data, 1 response is removed for being outside mainland U.S. (Hawaii) and 7 responses are removed due to being low-quality (straightlining, extreme hastiness, etc.), resulting in 691 usable data points.

Looking at the usable data, responses have been collected from 46 out of 48 states within mainland U.S. as well as Washington, D.C. The number of responses from the four most represented states and other sample statistics are shown in TABLE 1. The two unrepresented states, Vermont and Wyoming, are the least populated states, with 0.2% of the population each.

**TABLE 1 Sample statistics**

| Statistics[†] | Sample (responses) | Sample (%) | U.S. population/ other sources[*] (%) |
|---|---|---|---|
| **State** | | | |
| California | 110 | 15.9% | 12.0% |
| Texas | 55 | 7.9% | 8.8% |
| New York | 52 | 7.5% | 5.9% |
| Florida | 46 | 6.6% | 6.5% |
| **Gender** | | | |
| Male | 346 | 50.1% | 49.2% |
| Female | 331 | 47.9% | 50.8% |
| Non-Binary | 14 | 2.0% | - |
| **Age** | | | |
| 18-24 years | 207 | 30.0% | 9.3% |
| 25-34 years | 229 | 33.1% | 13.9% |
| 35-44 years | 136 | 19.7% | 12.8% |
| 45-54 years | 56 | 8.1% | 12.4% |
| 55-64 years | 39 | 5.6% | 12.9% |
| 65-74 years | 22 | 3.2% | 9.6% |
| 75 years or older | 2 | 0.3% | 6.8% |
| **Political Leaning** | | | |
| Democrat | 406 | 60.0% | 51.1% |
| Republican | 98 | 14.5% | 41.5% |
| Independent/no preference | 155 | 22.9% | 7.4% |
| Other (progressive, liberal, libertarian, …) | 18 | 2.6% | - |
| **Income** | | | |
| < $25,000 | 112 | 16.6% | 19.2% |
| $25,000 - $49,999 | 156 | 23.1% | 21.2% |
| $50,000 - $99,999 | 254 | 37.7% | 29.9% |
| $100,000 - $149,999 | 97 | 14.4% | 15.1% |
| $\geq$ $150,000 | 55 | 8.2% | 14.5% |

*Sources: state, gender and age (73); political leaning (74); income (75)

[†]Non-responses per category: state = 0, gender = 14, age = 0, political leaning = 14, income = 17

Looking at the age distribution, most of the sample (82.8%) are between the ages of 18 and 45, with mean and median age of 33.4 and 30 years respectively, resulting in a sample leaning towards younger adults. Politically, the sample is democrat leaning. Excluding respondents who preferred not to answer and other responses, 61.6% of the sample lean Democrat, 14.9% lean Republican and 23.5% consider their political views as independent or have no preference. While data has been collected in a slightly different manner, according to samples surveyed by Pew Research in 2018 (*74*), the latter percentages are 51.1%, 41.5% and 7.4% respectively. This bias towards Left-leaning partisanship is not, however, attributed to the higher response rates from states that tend to vote Democrat in recent elections, an observation confirmed by simulating elections using the sample distribution and election data (*76*). Instead, this bias is likely the result of self-selection in non-random online surveys (*77-79*). Finally, the average and median income for the sample are $72,100 and $62,500 annually, compared to $88,600 and $62,800 in the 2019 5-year American Community Survey (*75*). Accounting for the household size, the average and median annual household income for the sample are approximately $29,300 and $21,900 annually per household-member, respectively.

As for the response patterns of the likelihood of using hailed services before and during the COVID pandemic, the average response score shifts from 2.81 (closest to the scale midpoint of *neither likely nor unlikely*) to 2.18 (closest to *unlikely*). This means that as expected, on average we observe a decrease in the stated likelihood to use hailed modes during the pandemic (significant at α = 0.001 level). The distribution of responses is presented in TABLE 2. The shift in sentiment is further evidenced by the shift in (statistical) mode from 4 (*likely*) to 1 (*very unlikely*) as shown in Figure 1.

**TABLE 2 Percent distribution matrix for the likelihood of using hailed services before and during the COVID-19 pandemic (n = 691)**

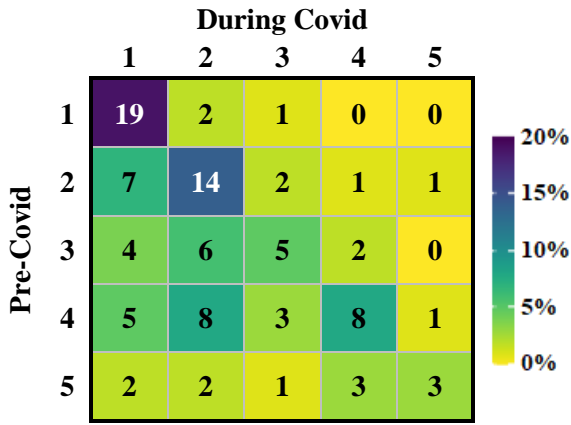

Nonetheless, as revealed by Figure 1 for the individual shift in likelihood of using hailed modes (calculated as the difference in likelihood of using hailed services *during covid* versus *before covid*), the largest share of respondents (48.5%) do not declare any change in behavior. The second largest group of respondents (42.1%) claim a decreased likelihood of use, countered by a small portion of respondents (9.4%) whose likelihood shifts positively. The uneven distribution of responses is important to note, as it will guide the modeling process and the importance of accounting for the ordinal nature of the data to properly capture the non-linearity in likelihood differences. Additionally, it is important to acknowledge that the same difference can be the result of two different starting points, potentially dictating different inherent behavioral meanings. These concerns are addressed in the ordinal regression model developed.

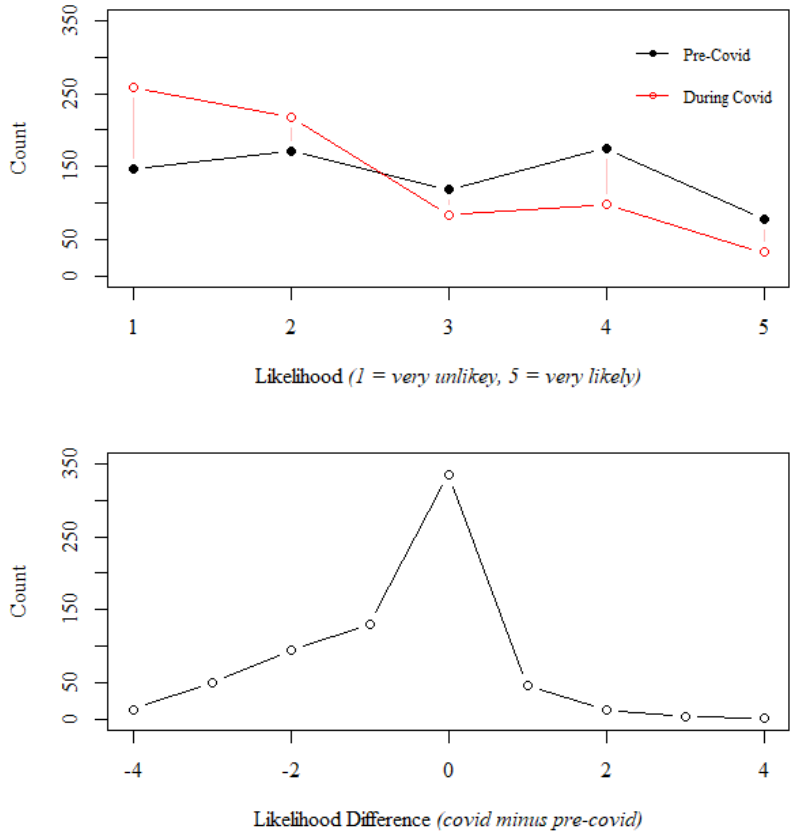

**Figure 1 Reported likelihood of using hailed services before and during COVID-19 pandemic (upper panel) and difference (lower panel)**

## METHODOLOGY AND STATISTICAL METHODS

We conduct a regression analysis to understand the relationship between several explanatory variables, such as demographics, typical travel behavior and political views, on the change in consideration of pandemic era on-demand rides. The dependent variable is defined by taking the difference between stated hailing likelihood before versus in the context of the COVID-19 pandemic. However, as discussed in the previous section and shown in Figure 1, the change in likelihood of using hailed modes is ordered in nature and distinctly non-linear. To account for ordinality and non-linearity, a *proportional odds model* is utilized. Additionally, for the environment- and technology-related latent attitudes, they are incorporated in the regression models serially through scores estimated as part of *confirmatory factor analysis*. Modeling details and approach are explained and discussed below.

### Ordinal Regression Model

The use of ordinal regression modeling allows us to account for the non-linear and ranked nature of responses within the modeling process, improving the forecasting accuracy of the resulting models (*80*). The ordinality of the dependent variable is accounted for through a latent continuous variable $y_i^*$ which is defined through a censoring approach as follows,

$$y_i = \begin{cases} y_1 & if & y_i^* \leq \mu_1 \\ y_2 & if & \mu_1 < y_i^* \leq \mu_2 \\ \vdots & \vdots & \vdots \\ y_{J-1} & if & \mu_{J-1} < y_i^* \leq \mu_J \\ y_J & if & \mu_J < y_i^* \end{cases} \qquad (1)$$

$y_i$ is the ranked (or ordinal) choice observed, $J$ is the number of ordered ranks (or stages), and $\mu_1$ to $\mu_J$ are a vector of threshold parameters to be estimated.

It is also assumed that the latent variable $y_i^*$ is unbounded, with $\mu_0$ being $-\infty$ and $\mu_{J+1}$ being $+\infty$. The resulting latent regression model has the familiar structure,

$$y_i^* = \boldsymbol{\beta}' \boldsymbol{x}_i + \varepsilon_i, \text{ where } i = 1, \dots, n \qquad (2)$$

where $n$ is the total number of individuals, $\boldsymbol{\beta}$ is a vector of coefficients, $\boldsymbol{x}_i$ are independent variables and $\varepsilon_i$ is the error term of a specified distribution, typically either logistic or normal.

The resulting probability on which the log-likelihood is estimated is presented below,

$$P(y_i = j | \boldsymbol{x}_i) = F(\mu_j - \boldsymbol{\beta}' \boldsymbol{x}_i) - F(\mu_{j-1} - \boldsymbol{\beta}' \boldsymbol{x}_i), \text{ where } j = 0, 1, \dots, J \qquad (3)$$

Here, $F$ is the selected cumulative distribution function (CDF), typically either logistic for an ordered logit – otherwise known as *proportional odds model* – or normal for an ordered probit model. $\mu_{j-1}$ is restricted to be less than $\mu_j$ to ensure that the above probability is positive for every $j$.

For more detailed treatment of ordered models, we refer readers to Greene and Hensher (*80*)(*79*). The resulting model, later presented in this paper, is estimated using *MASS v7.3.51.6* package (*81*) in *R v4.0.0 (82)* using a logistic error term.

**Ordinal Confirmatory Factor Analysis for Latent Variables**
The authors hypothesize that being tech-savvy results in a lower shift in likelihood to use hailed rides during the pandemic, if any. This is mainly due to the higher share of ridesharing among respondents and the dependence of these services on technology for requesting rides. Similarly, individuals who are more concerned about the environment may be more prone towards using hailed rides given the perceived complementarity between public transportation and ridesharing – even if this perception remains contested (*2, 31, 44, 83*).

Accordingly, two constructs are estimated, *tech-savviness* and *pro-environmentalism*, through exploratory and confirmatory factor analysis using *lavaan v0.6.6* package (*84*) in *R v4.0.0*. Specifically, the model is estimated using a diagonally weighted least squares estimator with robust standard errors and mean and variance adjusted test statistic, with restricted latent variable variances (for identification) and standardized indicators. The results and selected indicators are shown in TABLE 3. Note that some indicators may not be directly related to travel mode selection but are still viable indicators of general attitudes. Additionally, estimated threshold parameters are omitted for brevity, given that estimated threshold parameters do not have a direct interpretation but are mainly needed for estimating the cut-points in the model (*80, 85*).

**TABLE 3 Confirmatory factor analysis model for latent attitudes[+]**

| *Robust Model Statistics:* | *Chi-squared test statistic: 86.099* | *Comparative Fit Index (CFI): 0.972* | | |
|---|---|---|---|---|
| | *Degrees of Freedom: 19* | *Tucker-Lewis Index (TLI): 0.958* | | |
| | *Root Mean Square Error of Approximation (RMSEA): 0.072* | | | |
| Indicator | | Estimate | z-value | p-value |
| **Tech-Savviness** | | | | |
|   Technology is changing society for the better. | | 0.740*** | 27.78 | 0.000 |
|   I am excited to learn about new technologies in the market. | | 0.916*** | 43.81 | 0.000 |
|   I pay more to get more technologically advanced products. | | 0.694*** | 26.32 | 0.000 |
|   I use the internet daily for chatting and entertainment. | | 0.502*** | 12.29 | 0.000 |
| **Pro-Environmentalism** | | | | |
|   I am willing to switch to active mode of transportation (such as walking or cycling) in order to protect the environment. | | 0.732*** | 27.46 | 0.000 |
|   I would select more environmentally friendly package delivery options at the cost of slower delivery. | | 0.814*** | 34.03 | 0.000 |
|   I prefer to order items online in bulk to minimize the total number of delivery trips made to my address. | | 0.462*** | 12.67 | 0.000 |
|   I am concerned with the news about climate change. | | 0.705*** | 24.18 | 0.000 |
| **Covariance** | | | | |
|   $\rho_{(tech\text{-}savviness,\ pro\text{-}environmentalism)}$ | | 0.286*** | 6.295 | 0.000 |

*** = significant at 0.01 level; ** = significant at 0.05 level; * = significant at 0.10 level
[+] *Ordinal threshold effects are estimated but omitted from table for concision*

In terms of fit, according to CFI and TLI measures, the model provides a good fit (> 0.95 for both), although RMSEA should ideally be under 0.06 (*86*). All indicators are highly significant (α = 0.001) and have a positive sign in line with expectations. The two latent attitudes are positively correlated, suggesting that individuals who are more tech-savvy are also likely to be more pro-environment and vice-versa. Lastly, factor scores are calculated using the *Empirical Bayes Modal* method for each individual to be used serially in the regression model. For more information on factor analysis and structural equation modeling, the reader is referred to Bollen (*87*).

## RESULTS AND DISCUSSION

To capture the shift in hailing ridership intentions, a proportional odds model for the difference in likelihood to use hailed services <u>during</u> and <u>before</u> the pandemic is estimated. The dependent variable is the shift in user intentions during the pandemic, and five types of explanatory variables are controlled for to explain the change in consideration.

Nonetheless, from a modeling point of view, an important challenge related to building the model around the shift in likelihood (denoted by Δlikelihood) is the magnitude and directionality of potential shifts. The magnitude is affected by the starting likelihood values of individual respondents. This starting point is important to control for given that heterogeneity in pre-pandemic use of hailed services can shape current acceptability. Moreover, looking jointly at starting values and magnitudes of change, consider individuals whose initial likelihood is equal to 2 (*unlikely*). These individuals can only have a truncated set of Δlikelihood values within the set [-1, 0, 1, 2, 3]. On the other hand, individuals with an initial likelihood equal to 5 (*very likely*) can only have potential Δlikelihood values within the set [-4, -3, -2, -1, 0]. This bias due to initial likelihood and the respective set truncation is controlled for in the model by including ordered categorical variables for the initial likelihood values (before the pandemic) as independent variables, effectively transforming the model into a level-specific model (analogous to entity-specific models in econometrics (*88*).

**TABLE 4 Ordered regression model for the shift in consideration of hailed ride modes during the COVID-19 pandemic**

| Model Statistics: | Number of Observations: 691 | $\rho^2$: 0.156 | | |
|---|---|---|---|---|
| | Log-Likelihood at zero: -1044.99 | AIC: 1813.32 | | |
| | Final Log-Likelihood: -881.67 | BIC: 1926.79 | | |
| Parameter | | Coefficient | t-value | p-value |
| **Socioeconomic and Demographics** | | | | |
| UnderFortyFive: Under 45 years old (vs. 45 or older) | | -0.653*** | -3.000 | 0.003 |
| Male: Gender is male (vs. female and non-binary) | | 0.267* | 1.756 | 0.080 |
| Urban: Urban household (vs. suburban and rural) | | 0.474*** | 2.686 | 0.007 |
| IncomePerHH: Income per household (in $1,000) | | -0.00752** | -2.445 | 0.015 |
| MissingIncome: Income is not reported (vs. reported) | | -0.309 | -0.682 | 0.495 |
| **Typical Modes of Travel** | | | | |
| PrivateMode: Typical mode of travel is private vehicles (car and motorcycle) (vs. ridesharing, walking, cycling and taxi) | | -0.934*** | -3.327 | 0.001 |
| PublicMode: Typical mode of travel is transit (vs. ridesharing, walking, cycling and taxi) | | -0.502** | -2.403 | 0.017 |
| **Political Views** | | | | |
| DemLeft: Political view is either leaning Democrat, Democrat or other Left view (vs. independent, leaning Republican or other Right view) | | -0.465*** | -2.803 | 0.005 |
| PolViewMissing: Political view is not reported (vs. reported) | | -0.149 | -0.270 | 0.788 |
| **Latent Variable** | | | | |
| Pro-Environmentalism | | 0.212** | 2.290 | 0.022 |
| Tech-Savviness | | *n.s.* | - | - |
| **Impact of COVID** | | | | |
| StayAtHomeDuration: Length of stay-at-home orders for individual | | -0.183* | -1.757 | 0.079 |
| Quarantined: One or more household members have been quarantined because of COVID-19 | | 1.509*** | 3.090 | 0.002 |
| Quarantined x PrivateMode | | -1.236** | -2.275 | 0.023 |
| **Level-Specific Constants** | | | | |
| Previous likelihood of using hailed ride services = | | | | |
| *Very Unlikely* | | 2.056*** | 7.845 | 0.000 |
| *Unlikely* | | 1.306*** | 5.496 | 0.000 |
| *Likely* | | -0.971*** | -4.307 | 0.000 |
| *Very Likely* | | -1.904*** | -6.371 | 0.000 |
| **Thresholds of Consideration Shift** | | | | |
| $\mu_{(-4|-3)}$ | | -7.003*** | -12.39 | 0.000 |
| $\mu_{(-3|-2)}$ | | -5.191*** | -10.38 | 0.000 |
| $\mu_{(-2|-1)}$ | | -3.797*** | -7.974 | 0.000 |
| $\mu_{(-1|\ 0)}$ | | -2.515*** | -5.413 | 0.000 |
| $\mu_{(\ 0|+1)}$ | | 0.905* | 1.942 | 0.053 |
| $\mu_{(+1|+2)}$ | | 2.304*** | 4.603 | 0.000 |
| $\mu_{(+2|+3)}$ | | 3.515*** | 5.839 | 0.000 |
| $\mu_{(+3|+4)}$ | | 4.628*** | 5.548 | 0.000 |

*** = significant at 0.01 level; ** = significant at 0.05 level; * = significant at 0.10 level

Controlling for these aspects, the resulting model is shown in TABLE 4 with threshold effects. Unless stated otherwise, coefficients that are considered significant are <u>significant at the 0.05 level</u>.

The model identifies a number of factors impacting the change in consideration. We note that many of the findings about shared travel during the pandemic contradict the typical ridership profiles found in the ridehailing literature. These unexpected findings for age, gender and income will be carefully contextualized in the following subsection.

**Demographic Factors**

Starting with socio-demographics, individuals below 45 years have a significant and negative coefficient, indicating a decreased willingness to use on-demand rides during the pandemic. While seemingly counterintuitive, given the evidence that younger individuals more readily adopt ridehailing (*13, 50, 89*), the decreasing interest is due to significant disparity in pre-pandemic use of ridehailing modes between younger and older individuals, as shown in Figure 2. Given the higher pre-pandemic demand, and likelihood that young respondents are engaged with more remote-friendly education and work requirements, we observe a larger drop in likelihood of using these services during the pandemic for younger adults.

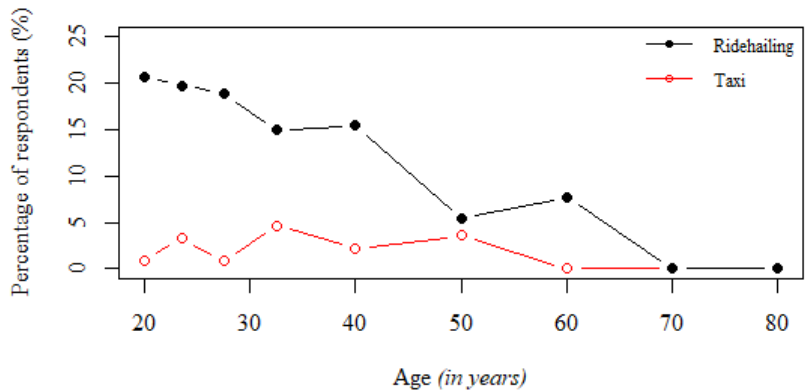

**Figure 2 Percentage (by age category) of hailed-ride service users prior to the pandemic as a function of age**

While gender is only significant at the 0.10 level, it suggests that men are more prone to consider hailed services in comparison to women and non-binary individuals. This finding confirms the observation by Lavieri and Bhat (*12*) that women are less likely to use ridehailing for routine commitments, such as family and shopping errands, which are likely to fall under the category of essential trips. The gender differences presumably stem from higher concerns regarding security and safety among women towards hailed modes, be it taxi or ridehailing (*2, 90*). We note that these concerns may be amplified during the early stages of the pandemic where errand trips are fewer and more critical and where public areas are less crowded and isolating. While transportation literature on non-binary individuals is limited (*2*), some articles report that experiences of harassment and discrimination on transit and ridesharing – are not dissimilar to concerns observed among women in some instances (*91, 92*).

Urban dwellers, compared to respondents living in suburban or rural areas, have a significant increase in their consideration of hailed modes during the pandemic. This may reflect, especially for rural regions, the lower adoption levels of hailed services (*93*). Non-urban dwellers are less accustomed to these services and may perceive them as posing higher risk for infection. Additionally, in response to the pandemic and loss of ridership, there has been a reduction in active taxi and ridehailing drivers, resulting in undesirable waiting times and poor service, particularly outside of dense areas (*24, 94*).

Higher household income has a significant negative impact on Δlikelihood. In typical conditions, this result may stand out and even be considered counterintuitive as higher income is often associated with increased use of hailed services (*10, 95*). Nonetheless, early reports and studies have shown that higher income individuals and households have more promptly reacted and adapted to the risks of exposure to COVID-19. Specifically, higher income households have, on average, a higher ability to adapt their lifestyles to the pandemic owing to more secure work, resources for homeschooling, higher prospect of being able to work from home and ability to reduce travel as well as eliminate non-essential trips (*56, 57, 96-98*). The resulting decrease in consideration by individuals from higher income households is, therefore, reflective of a more general mobility trend among that demographic. Missing income values are controlled for through the dummy variable *MissingIncome* but reveals no significant behavioral difference for individuals who opt to not report their incomes, as indicated by the insignificant coefficient.

**Travel Mode Habits**
It is essential to closely examine the changing relationships between modes, as the pandemic era is fueling massive shifts in ridership and possibly lasting changes in perceptions of which modes can best meet evolving travel needs. Not unexpectedly, individuals who typically travel by private modes (incl. motorcycles) tend to strongly decrease their use of ridehailing or taxi during the pandemic. It is intuitive that individuals who have more access to and familiarity with private options will gravitate towards these modes during the precarious pandemic period. This effect is even more pronounced for individuals who have been quarantined during the pandemic, as implied by the interaction term. This suggests that experiencing a more severe mobility restriction due to viral exposure will lead to even more reluctance towards sharing.

The relation between public transportation and hailing-based shared modes is more complex. It appears likely that the experience of navigating public transit could lead to more acceptance of hailing a vehicle during the pandemic restrictions as both models imply sharing, albeit occurring in a more confined setting. Instead, we find that transit users (as compared to active and hailing travelers) would reduce their use of hailed services. Yet, the difference in coefficient magnitude indicates that the aversion to sharing is less strong in the transit user segment compared to the private mode users. We speculate therefore that in terms of ranking the perceived risk of using different transportation modes, transit is ranked 'the middle option' between private modes and ridehailing. This reflects opinions shared by Uber's Director of Policy, Shin-pei Tsay, in a July 2020 seminar hosted by Northwestern University Transportation Center (*98*).

**Attitudes and Political Orientation**
Political views have a significant impact on the shift in consideration during the pandemic. In particular, individuals who reported their political orientation as either *Democratic* or *Left-leaning* (e.g. *Progressive, Socialist*, etc.) have a negative Δlikelihood (in comparison to those with *Independent* and *Republican* political views). This result suggests, in line with recent reports (*99-103*), an increased level of concern about the pandemic and risk-averse behavior from individuals with more progressive political views. Fourteen (2.0%) respondents opted not to report their political views; however, these respondents show no behavioral differences as implied by the insignificant coefficient for *PolViewMissing*.

The latent indicator *Pro-Environmentalism* has a positive and significant coefficient. While there is still on-going debate on the extent to which ridehailing benefits or harms the environment (*83*), the general perception and operator marketing of these modes tend to suggest they are environmentally sustainable alternatives (*2*), especially through the consideration of pooled rides. This sentiment may be the driving force for this increased consideration of hailed modes by individuals who are more cognizant about the environment. The latent variable for tech-savviness has been considered in the model and a moderately positive effect has been observed, indicating that pro-technology attitudes are associated with a shift towards hailed rides (likely mainly driven by ridehailing). However, the parameter was insignificant in the final version and, therefore, removed.

**COVID-19 Impacts**

Finally, looking at variables directly related to the impact of COVID on individuals' lifestyles, individuals living in areas or states that have been under *stay-at-home* orders for a longer period of time have lower consideration of hailed modes during the pandemic (at a lower significance of 0.10). At the time of data collection (June 2020), the length of *stay-at-home* orders has been reflective of both the severity of the virus impact within an area, as well as the general vigilance towards the associated risks. For example, San Francisco, one of the areas significantly impacted by the virus early on in the pandemic, was the first region within the U.S. to impose stay-at-home restrictions (*101*). Conversely, Florida, one of the later states to adopt social distancing measures, has had multiple reports of the public negligence towards these measures (e.g. (*102*)).

Interestingly, having been under quarantine has the opposite sign (positive) to spending more time under a *stay-at-home order*. We speculate that the more severe restriction on travel resulting from quarantining, which typically lasts for 14 days, causes individuals to compensate by expanding mobility via shared ride modes. For individuals who typically travel by private modes this positive effect is largely canceled out, down to 0.273 from 1.509 ($\beta_{Quarantined} + \beta_{Quarantined\ x\ PrivateMode} = 0.273$). Not unexpectedly, the extent to which individuals need to compensate for restricted mobility due to quarantining is more muted for individuals who typically travel by private mode.

**Other Model Observations**

Finally, the level-specific constants, as discussed, are constants that both, capture the innate characteristics of respondents not controlled for in the model and control for the statistical and status-quo biases in the dependent variable. As for the threshold estimates, they are considered "nuisance parameters", with limited value for interpretation but rather necessary for estimation (*80, 85*).

Other important variables that were tested but ultimately not found to be significant in the final model are *ethnicity* and *employment*. Specifically, during the modeling process, a positive relationship between Black ethnicity and hailing consideration has been observed. For employment, individuals who are *unable to work* also have been observed to have increased consideration of hailing. The latter may be a proxy for health or physical limitations. It is hypothesized that COVID-19 disproportionality impacts individuals with lower mobility, therefore making any viable source of increased mobility desirable.

**CONCLUSION**

This paper presents an analysis of the shift in consideration of shared on-demand mobility in the U.S. as a result of the COVID-19 pandemic. Specifically, the modes examined in this study is that of ridehailing, such as *Uber* and *Lyft*, and taxi services. The paper has a number of findings that are relevant to the research and planning communities alike, revealing novel insights on the role of political belief and pandemic experiences for the intention to use hailed services. Essentially, this study reveals the inner motivations of who increases, decreases or retains their intended use of ridesharing during the pandemic.

The model confirms several research findings on ridehailing patronage in ordinary (pre-pandemic) contexts, concerning significant impact of gender, pro-environmental views and living in an urban setting. Particular attention, however, is warranted to some other parameters in the model. A decrease in hailed ride consideration is expected among younger cohorts, the main adopters of ridehailing services under normal operations. Additionally, while there is extra complexity in navigating public transit in comparison to hailed services, transit riders gravitate away from hailed rides during the pandemic. With increasing income, consideration of hailed modes is also reduced throughout the pandemic, potentially as a result of having greater control over lifestyle, mobility and viral exposure among higher income individuals.

The effect of political views on the consideration of hailed mobility is especially interesting. This study finds that individuals with Democratic or Left-leaning political views tend to have lower consideration for hailed rides as a result of the pandemic. With the developing political climate surrounding COVID-19 and consequent measures, insight on relationships between partisan leaning and risk-perceptions, vigilance and mobility behavior during an ongoing health crisis is still a developing topic. This paper suggests, along with other early observations (*100, 103-105*), that attitudes and behavior towards suggested or enforced COVID measures are impacted by political leaning. In our study this holds even after controlling for income. Finally, when controlling for pandemic experiences, contrary to expectation, individuals in households that have had a quarantined member, increased their consideration of using hailed modes. It is hypothesized that this increased consideration is a compensation mechanism as a result of the strenuous mobility restrictions resulting from a quarantine.

All in all, this study presents novel findings on the consideration of different mobility solutions during a health emergency with increased complexity in decision making, caused by mobility restrictions and concerns of disease transmission. The results aid in understanding the shift in consideration of hailed ride services as a result of the COVID-19 pandemic, and sheds light on important variations among different user segments. This research contributes new insight on how mobility decision-making and use intentions are altered during the pandemic and helps pinpoint important determinants of mobility reductions. Going forward, these determinants are likely to play an important role in shaping decisions around shared rides and modes and helps anticipate the behavioral impacts of drastic changes in lifestyle on transportation systems use during pandemic outbreaks and aftermaths.

**Limitations and Future Work**
One limitation of this study is the use of cross-sectional data to measure change in consideration over time. Ideally, the analysis would have used longitudinal data collected at two points in time, before the pandemic and during. Nonetheless, given the unpredictable nature and severity of the pandemic, acquiring data in such a longitudinal manner is challenging. While the authors acknowledge the biases inherent in asking respondents to recall their past consideration of using a mode, the data still offers valuable insight into the impact of the pandemic on this consideration process. Additionally, this study uses convenience online sampling recruitment (first-come, first-serve basis), resulting in a sample that is more representative of politically Left-leaning, lower-income and younger respondents. As part of the modeling process – and to reduce this bias, these sociodemographic features are controlled for in the model presented in this study. Another limitation stems from survey length constraints and early timing in the pandemic process, where only a core set of pandemic questions were included. Future research linking (shared) mobility behavior to pandemic effects should include a more exhaustive list of questions directly relevant to COVID-19, such as mask-wearing habits, vaccination behavior, pandemic fatigue, and sources of opinion formation surrounding the pandemic, to better parse how belief is founded and affecting mobility behavior. Another potential extension for this study is estimating the measurement equations for latent attitudes simultaneously with the ordinal regression model. In particular, partisan views or ability to control pandemic exposure is likely to affect and be affected by attitudes, that in turn affect behavior. A more detailed structural equation model can aid our understanding of causation among these factors and improve model efficiency and accuracy.

## ACKNOWLEDGMENTS

This research was partially supported by Northwestern Transportation Center Terminal Year Fellowship and by the US National Science Foundation Career Award No. 1847537. The authors would like to thank Elisa Borowski, Emma Zajdela, Hoseb Abkarian, Nour El Assi and other survey pilots for their early feedback on the distributed survey. The authors would also like to thank the anonymous reviewers for their time, input and recommendations. The study was approved by Northwestern's Institutional Review Board under study number *STU00212452*.

## AUTHOR CONTRIBUTIONS

The authors confirm contribution to the paper as follows: study conception and design: M. Said and A. Stathopoulos; data collection: M. Said and A. Stathopoulos; analysis and interpretation of results: M. Said and A. Stathopoulos; draft manuscript preparation: M. Said, J. Soria and A. Stathopoulos. All authors reviewed the results and approved the final version of the manuscript.